\documentclass[prd,preprint,showpacs]{revtex4}
\usepackage{amsmath}
\usepackage{graphicx}
\usepackage{bm}
\begin{document}
\title{Relativistic ideal Fermi gas at zero temperature and preferred frame}
\author{K. Kowalski, J. Rembieli\'nski and K.A. Smoli\'nski}
\affiliation{Department of Theoretical Physics, University
of \L\'od\'z, ul.\ Pomorska 149/153, 90-236 \L\'od\'z,
Poland}
\begin{abstract}
We discuss the limit $T\to 0$ of the relativistic ideal Fermi gas of
luxons (particles moving with the speed of light) and tachyons 
(hypothetical particles faster than light) based on observations of
our recent paper: K. Kowalski, J. Rembieli\'nski and K.A. Smoli\'nski,
Phys. Rev. D {\bf 76}, 045018 (2007).  For bradyons this limit is in 
fact the nonrelativistic one and therefore it is not studied herein.
\end{abstract}
\maketitle
\section*{}
In our very recent paper \cite{1} the Lorentz covariant formulation
has been introduced of classical and quantum statistical mechanics
and thermodynamics of an ideal gas of relativistic particles.  An
advantage of this formulation based on the preferred frame approach
is among others the possibility of a consistent free of paradoxes
description of tachyons.  In our discussion of the limit $T\to 0$ in
the case of the relativistic quantum ideal gas we restricted in
\cite{1} to the case of the Bose gas.  In this work we complete the
results given in \cite{1} by considering the limit $T\to 0$ for the
Fermi gas.  As with the Bose gas, we do not discuss the limit $T\to
0$ in the case of bradyons (particles slower than light) because
this limit is in fact the nonrelativistic one \cite{2}.  Consider
first the degenerate Fermi gas of tachyons.  The covariant forms of
thermodynamic functions for that gas derived in \cite{1} are given by
\begin{eqnarray}
%<1>
\frac{U}{V} &=&
\frac{m^2}{2\pi^2u_0^2\beta^2}\sum_{n=1}^{\infty}\frac{
(-1)^{n+1}}{n^2}[3S_{0,2}(nu_0\beta m) - nu_0\beta mS_{-1,1}(nu_0\beta
m)]s^n,\\
\frac{p}{kT} &=& \frac{m^2}{2\pi^2u_0^2\beta}\sum_{n=1}^{\infty}
\frac{(-1)^{n+1}}{n^2}S_{0,2}(nu_0\beta m)s^n,\\
\frac{N}{V} &=& \frac{m^2}{2\pi^2u_0^2\beta}\sum_{n=1}^{\infty}\frac{
(-1)^{n+1}}{n}S_{0,2}(nu_0\beta m)s^n,
\end{eqnarray}
where $\beta=1/(kT)$, $m$ is a mass of a particle, $s=e^{\beta\mu}$ is the 
fugacity, $\mu$ is the chemical potential, $u_0$ is the zeroth (covariant) 
component of the four velocity $u^\mu$ of the preferred frame with repect 
to the inertial observer (in the preferred frame $u_0=1$), and
$S_{\mu,\nu}(x)$ is the Lommel function (see the appendix).  Using
(A.5) we obtain the following asymptotic form of (1), (2), and (3) in the 
limit $T\to0$ (i.e.\ $\beta\to\infty$):
\begin{eqnarray}
%<4>
\frac{U}{V} &=& \frac{m}{\pi^2u_0^3\beta^3}\sum_{n=1}^{\infty}
\frac{(-1)^{n+1}}{n^3}s^n,\\
\frac{p}{kT} &=& \frac{m}{2\pi^2u_0^3\beta^2}\sum_{n=1}^{\infty}
\frac{(-1)^{n+1}}{n^3}s^n,\\
\frac{N}{V} &=& \frac{m}{2\pi^2u_0^3\beta^2}\sum_{n=1}^{\infty}
\frac{(-1)^{n+1}}{n^2}s^n.
\end{eqnarray}
Now applying the asymptotic form of the Fermi functions
\begin{equation}
%<7>
f_\nu(x) = \sum_{n=1}^{\infty}\frac{(-1)^{n+1}}{n^\nu}x^n
\end{equation}
known as Sommerfeld's lemma \cite{3}, such that
\begin{eqnarray}
%<8>
&&f_\nu(e^{\beta\mu})\\\nonumber
&& \simeq \frac{(\beta \mu)^\nu}{\Gamma(\nu+1)}
\left[1+\nu(\nu-1)\frac{\pi^2}{6}(\beta\mu)^{-2}+\nu(\nu-1)(\nu-2)
(\nu-3)\frac{7\pi^4}{360}(\beta \mu)^{-4}+\ldots\right],\quad 
\beta\mu \gg 1,
\end{eqnarray}
we find
\begin{eqnarray}
%<9>
\frac{U}{V} &=&
\frac{m\mu^3}{6\pi^2u_0^3}\left(1+\pi^2\frac{k^2T^2}{\mu^2}\right),\\
p &=& \frac{m\mu^3}{12\pi^2u_0^3}\left(1+\pi^2\frac{k^2T^2}{\mu^2}\right),\\
\frac{N}{V} &=& \frac{m\mu^2}{4\pi^2u_0^3}\left(1+\frac{\pi^2}{3}\frac{k^2T^2}
{\mu^2}\right).
\end{eqnarray}
We remark that as with the case of the Bose gas of tachyons \cite{1}, (9) and (10)
imply the equation of state of the form
\begin{equation}
%<12>
pV = \frac{U}{2}.
\end{equation}
Now, it follows immediately from (11) that the Fermi energy $\varepsilon_{\rm F}$ 
is given by
\begin{equation}
%<13>
\varepsilon_{\rm F} = \sqrt{\frac{4\pi^2u_0^3N}{mV}},
\end{equation}
where $\varepsilon_{\rm F}=\mu(T=0)$.  Furthermore, eqs.\ (11) and
(13) yield
\begin{equation}
%<14>
\varepsilon_{\rm F}^2 = \mu^2+\frac{\pi^2}{3}k^2T^2,
\end{equation}
which leads to
\begin{equation}
%<15>
\mu = \varepsilon_{\rm F} -\frac{\pi^2}{6\varepsilon_{\rm F}}k^2T^2.
\end{equation}
We point out that in this section we perform calculations up to terms
of order $T^2$.  Making use of (14) and (15) we can write (9) and (10) 
in the following form:
\begin{eqnarray}
%<16>
U &=& \frac{mV}{6\pi^2u_0^3}\left(\varepsilon_{\rm F}^3 + \frac{\pi^2}{2}
\varepsilon_{\rm F}k^2T^2\right),\\
p &=& \frac{m}{12\pi^2u_0^3}\left(\varepsilon_{\rm F}^3 + \frac{\pi^2}{2}
\varepsilon_{\rm F}k^2T^2\right).
\end{eqnarray}
Consider now the entropy which can be defined as
\begin{equation}
%<18>
S = \frac{U+pV-\mu N}{T}.
\end{equation}
Taking into account (9), (10), (11) and (15) we get
\begin{equation}
%<19>
S = \frac{mk^2}{6u_0^3}V\varepsilon_{\rm F}T.
\end{equation}
Therefore, the entropy vanishes at $T=0$.  The formula (19) can be
also obtained from the well-known relation
\begin{equation}
%<20>
S=\int_{0}^{T}\frac{C_V}{T}dT,
\end{equation}
where
\begin{equation}
%<21>
C_V = \left(\frac{\partial U}{\partial T}\right)_V=\frac{mk^2}{6u_0^3}
V\varepsilon_{\rm F}T,
\end{equation}
which is immediate consequence of (16).

Finally, we study the limit $T\to0$ in the case of the Fermi gas of luxons.  
Using the formulas on thermodynamic functions derived in \cite{1} such that
\begin{eqnarray}
%<22>
\frac{U}{V} &=& \frac{3}{\pi^2u_0^4\beta^4}\sum_{n=1}^{\infty}
\frac{(-1)^{n+1}}{n^4}s^n,\\
\frac{p}{kT} &=& \frac{1}{\pi^2u_0^4\beta^3}\sum_{n=1}^{\infty}
\frac{(-1)^{n+1}}{n^4}s^n,\\
\frac{N}{V} &=& \frac{1}{\pi^2u_0^4\beta^3}\sum_{n=1}^{\infty}
\frac{(-1)^{n+1}}{n^3}s^n,
\end{eqnarray}
and proceeding analogously as with tachyons, we obtain
\begin{eqnarray}
%<25>
\frac{U}{V} &=&
\frac{\mu^4}{8\pi^2u_0^4}\left(1+2\pi^2\frac{k^2T^2}{\mu^2}\right),\\
p &=& \frac{\mu^4}{24\pi^2u_0^4}\left(1+2\pi^2\frac{k^2T^2}{\mu^2}\right),\\
\frac{N}{V} &=& \frac{\mu^3}{6\pi^2u_0^4}\left(1+\pi^2\frac{k^2T^2}
{\mu^2}\right).
\end{eqnarray}
We point out that (25) and (26) imply the well-known equation of state for
ideal gas of massless particless such that
\begin{equation}
%<28>
pV = \frac{U}{3}.
\end{equation}
Furthermore, the Fermi energy is
\begin{equation}
%<29>
\varepsilon_{\rm F} = \left(6\pi^2u_0^4\frac{N}{V}\right)^{\frac{1}{3}}.
\end{equation}
Hence, using (27) we find
\begin{equation}
%<30>
\varepsilon_{\rm F}^3 = \mu^3+\pi^2k^2T^2\mu,
\end{equation}
implying
\begin{equation}
%<31>
\mu = \varepsilon_{\rm F} -\frac{\pi^2}{3\varepsilon_{\rm F}}k^2T^2.
\end{equation}
The relations (25) and (26) written in terms of the Fermi energy
take the form
\begin{eqnarray}
%<32>
U &=& \frac{V}{8\pi^2u_0^4}\left(\varepsilon_{\rm F}^4 +
\frac{2\pi^2}{3}\varepsilon_{\rm F}^2k^2T^2\right),\\
p &=& \frac{1}{24\pi^2u_0^4}\left(\varepsilon_{\rm F}^4 +
\frac{2\pi^2}{3}\varepsilon_{\rm F}^2k^2T^2\right).
\end{eqnarray}
The entropy calculated with the help of (18) is
\begin{equation}
%<34>
S = \frac{k^2}{6u_0^4}V\varepsilon_{\rm F}^2T.
\end{equation}
Thus, as with tachyons, the entropy in the case of fermionic luxons
also vanishes.  The above formula on the entropy is also implied by
(20) and
\begin{equation}
%<35>
C_V = \left(\frac{\partial U}{\partial T}\right)_V=\frac{k^2}{6u_0^4}
V\varepsilon_{\rm F}^2T,
\end{equation}
following directly from (32).

In conclusion, we have derived in this work the Lorentz covariant form 
of thermodynamic functions of Fermi ideal gas of luxons and tachyons
in the zero temperature limit.  It seems that the observations of
this paper would be of interest for testing the hypothesis of
tachyonic neutrinos \cite{4} and tachyonic dark matter \cite{5}. Indeed, 
the existing estimations of density of neutrinos suggest that the model 
of an ideal gas is appropriate for the neutrino background.
\section*{Acknowledgements}
This paper has been supported by University of Lodz grant.
\appendix*
\section{}
We recall some basic properties of the Lommel functions
$S_{\mu,\nu}(x)$ \cite{6}.  The Lommel functions $S_{0,\nu}(x)$ have
the integral representation such that
\begin{equation}
%<a.1>
S_{0,\nu}(x) = \int_{0}^{\infty}e^{-x\sinh t}\cosh\nu t\,dt
= \frac{x}{\nu}\int_{0}^{\infty}e^{-x\sinh t}\cosh t\sinh\nu
t\,dt,\qquad x>0.
\end{equation}
The Lommel functions $S_{\nu,\nu}(x)$ can be expressed by means of
the Struve functions ${\bf H}_\nu(x)$ and the Bessel functions
$Y_\nu(x)$ (Neumann functions) also designated by $N_\nu(x)$.
Namely, we have
\begin{equation}
%<a.2>
S_{\nu,\nu}(x) = 2^{\nu-1}\sqrt{\pi}\Gamma(\nu + \hbox{$\scriptstyle
1\over 2$})[{\bf H}_\nu(x)-Y_\nu(x)].
\end{equation}
The Lommel functions $S_{\mu,\nu}(x)$ are defined recurrently
\begin{gather}
%<a.3>
S_{\mu+2,\nu}(x) = x^{\mu+1}-[(\mu+1)^2-\nu^2]S_{\mu,\nu}(x),\\
S'_{\mu,\nu}(x) = \frac{\nu}{x}S_{\mu,\nu}(x)+(\mu-\nu-1)S_{\mu
-1,\nu +1}(x) = -\frac{\nu}{x}S_{\mu,\nu}(x)+(\mu+\nu-1)S_{\mu
-1,\nu -1}(x).
\end{gather}
The asymptotics $S_{0,2}(x)$  for $x\gg 1$ is of the form
\begin{equation}
S_{0,2}(x) = \frac{1}{x},\qquad x\gg 1.
\end{equation}

\end{document}